\documentclass[aps,english,prb,twocolumn,superscriptaddress]{revtex4}
\usepackage{graphicx}
\makeatletter
\def\p@figure{Figure~}
\makeatother

\begin{document}

\title{Modeling spontaneous charge transfer at metal/organic hybrid heterostructures}

\author{V. Ongun \"{O}z\c{c}elik}\email{ongunozcelik@gmail.com}
\affiliation{Department of Chemistry and Biochemistry, University of California San Diego, La Jolla, CA 92093 USA}
\affiliation{Materials Science and Engineering Program, University of California San Diego, La Jolla, CA 92093 USA}

\author{Yingmin Li}
\affiliation{Materials Science and Engineering Program, University of California San Diego, La Jolla, CA 92093 USA}

\author{Wei Xiong}
\affiliation{Department of Chemistry and Biochemistry, University of California San Diego, La Jolla, CA 92093 USA}
\affiliation{Materials Science and Engineering Program, University of California San Diego, La Jolla, CA 92093 USA}

\author{Francesco Paesani}\email{fpaesani@ucsd.edu}
\affiliation{Department of Chemistry and Biochemistry, University of California San Diego, La Jolla, CA 92093 USA}
\affiliation{Materials Science and Engineering Program, University of California San Diego, La Jolla, CA 92093 USA}
\affiliation{San Diego Supercomputer Center, University of California San Diego, La Jolla, CA 92093 USA}

\begin{abstract}

Hybrid materials are crucial in photovoltaics where the overall efficiency of the heterostructure is closely related to the level of charge transfer at the interface. Here, using various metal / poly(3-hexylthiophene)(P3HT) heterostructure models, we reveal that the level of spontaneous charge transfer and electronic coupling at these interfaces depend on the conformational regularity of the organic polymer deposited on the metal substrate. Using ab-initio quantum chemical calculations  based on density functional theory (DFT) and heterodyne vibrational sum frequency generation (HD-VSFG) measurements, we show that inducing regio-randomness into the organic polymer modifies the intensity of interfacial electronic states, level of hybridization, density of interfacial charge transfer and the electronic wave function of the material. We present the HD-VSFG responses of the metal/P3HT heterojunctions containing both regio-regular and regio-random P3HT structures and show that the amount of non-resonant signal is closely related to the level of the spontaneous charge transfer at the interface. Thus, by measuring the non-resonant response of the metal/P3HT heterojunctions, the level of spontaneous charge transfer at the interface can be determined.

\end{abstract}

\maketitle

Integration of surfaces with ultra thin materials is essential for designing hybrid functional devices where the unique functionalities arising at their interface strongly influence the overall properties of the heterostructure beyond those of the isolated materials.\cite{bae2019integration} In this regard, the development of hybrid heterostructures comprising an inorganic substrate and a thin film of organic polymer is a recent milestone achieved in molecular electronics, battery electrodes  and photovoltaics research.\cite{huynh2002hybrid, yin2004colloidal, coakley2004conjugated, chang2010high} Upon creating a physical contact between a polymer and a metallic substrate, in the absence of an external bias voltage, charge separation can happen in the heterostructure where the hole states are localized in the polymer and the electrons transfer into the substrate, which we refer herein as spontaneous charge transfer. The spontaneous charge transfer between the metallic region and the organic polymer plays a crucial role in defining the optical and electronic efficiencies of the heterostructure since the charge accumulated around the contact region is likely to influence the interfacial electronic structure and coupling.\cite{braun2009energy} This process depends on the atomic structure of the junction, the chemistry of the substrate and the organic polymer, conformation of the polymer on the substrate and the level of hybridization between the substrate and the polymer.\cite{dag2008modeling, dag2010packing} Thus, a precise fundamental understanding of these properties and their effect on the spontaneous charge transfer mechanism is the first step toward the molecular engineering of metal / organic interfaces which will be used in macroscopic assembly of solid state structures for molecular electronics and conductive coating for battery electrodes.

In photovoltaic applications, using P3HT as the conjugated organic polymer in hybrid heterostructures is very common due to its high solubility and environmental stability.\cite{coakley2004conjugated,dang2011p3ht, holliday2016high} In particular, it has a remarkable electron mobility of 0.2 cm$^{2}$V$^{-1}$s$^{-1}$ which is one of the highest reported for an organic polymer where this high mobility is caused by the formation of lamellar structure of the backbone and the alkyl side chains present in P3HT.\cite{wang2004poly, sirringhaus1999two} However, in hybrid heterojunctions of conjugated organic polymers, the spontaneous charge transfer process between the substrate and the polymer is often difficult to monitor experimentally with surface scanning techniques such as scanning tunneling microscopy or atomic force microscopy due to the fact that electron transfer occurs at hidden interfaces.\cite{anglin2010monitoring, xiang2017ultrafast} Photoelectron spectroscopy is often applied to extract the band structure at interfaces, which is still limited to thin film interfaces due to the short penetration depth of photoelectrons.\cite{lindau1974probing}

\begin{figure*}
\includegraphics[width=15cm]{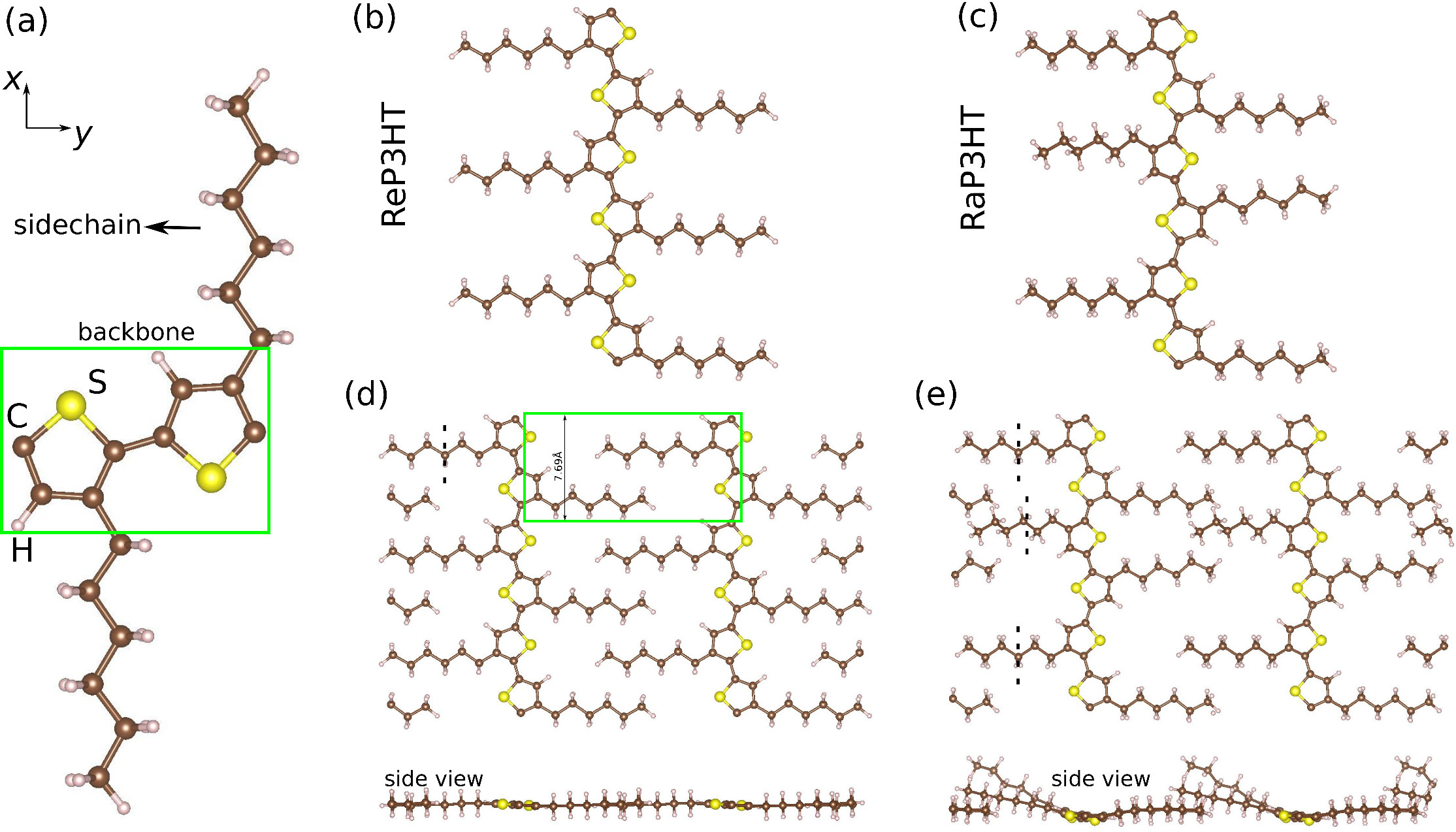}
\caption{Optimized geometries of various P3HT conformations. (a) P3HT monomer, (b) regioregular chain (ReP3HT), (c) regiorandom chain (RaP3HT) (d-e) planar allotropes of ReP3HT and RaP3HT chains. In the ball-and-stick model C, S and H atoms are represented with brown, yellow and pink spheres, respectively. }
\label{fig1}
\end{figure*}

In this letter, we use quantum chemical calculations based on DFT and experimental HD-VSFG measurements to provide insight on the nanoscale mechanisms of spontaneous charge transfer process in metal / P3HT interfaces. On the theoretical front, by calculating various stable conformations of P3HT (regio-regular, regio-random, planar, chain) on different metal substrates (Au, Ag, Pt) we reveal that (i) the level of charge transfer and electronic coupling at the interface can be selectively engineered by controlling the conformational regularity of the organic polymer, (ii) the geometrical regio-randomness of the organic polymer not only changes the charge transfer between the layers but it also determines how interlayer states arise at the interface, and (iii) these interlayer states significantly modify the level of hybridization, surface carrier density, local electrostatic potential and the wave function of the material. We experimentally monitor the hidden interfaces in these heterojunctions by measuring the non-resonant and resonant responses of their HD-VSFG signals  and show that the amount of non-resonant signal is closely related to the level of charge transfer at the interface. Thus, using both theoretical and experimental results, we reveal that HD-VSFG signal can be used as an indicator for probing spontaneous interfacial charge transfer and demonstrate that the spontaneous charge transfer between P3HT and metal substrate significantly depends on the regio-regularity of the polymer.

We start our analyses by calculating the optimized ground state structure of an isolated P3HT molecule that corresponds to the energy minimum on the Born-Oppenheimer surface. The geometry of the isolated P3HT molecule is presented in \ref{fig1}(a), where the monomer consists of a $\pi$-conjugated thiophene backbone and two alkyl side chains. P3HT monomers can be arranged in a sequential order to construct various types of P3HT polymer chains which differ by how the side chains are located in alternating monomers as shown \ref{fig1}(b-c). Accordingly, these polymers can be divided into two subgroups: (i) regio-regular (ReP3HT) polymers where the side chains are arranged in the same orientation in each monomer and (ii) regio-random (RaP3HT) polymers where the side chains are attached to the backbone from random locations in alternating monomers. In its ground state, the chain length in a period along the y-direction is 7.69\AA~ which is in excellent agreement with previous experimental result of 7.7\AA.\cite{sirringhaus2000microstructure} In ReP3HT, the axes of the side chains remain perpendicular to the axis of the backbone during conjugate gradient geometry relaxation steps, whereas the side chains rotate in RaP3HT. We note that, although there are various possible ways to construct RaP3HT, in \ref{fig1}(c) we show only one of those arrangements. Another allotrope of P3HT, shown in \ref{fig1}(d), is obtained by bringing individual periodic chains closer to each other on the same plane to create a thin layer of P3HT. In this geometry, the alkyl side chains of alternating P3HT units penetrate toward each other where the distance between the backbones of adjacent chains is 13.42\AA. Here it should be noted that as the chains are brought closer to each other, the ReP3HT structure preserves its initial planar geometry.

If we repeat a similar procedure to construct a thin layer of RaP3HT, the side chains rotate asymmetrically along backbone as shown in \ref{fig1}(e). This is primarily due to the fact that the minimum distance between the side chains of adjacent polymers is lower in the RaP3HT as compared to the ReP3HT. These rotations induce further interaction between the side chains. In other words, while the C-C bond distances within a side chain of the ReP3HT are identical, the final stable structure of the RaP3HT have non-uniform bond length distribution in its side chains. For ReP3HT, the geometrical midpoint of each side chain aligns with its middle carbon atom (as indicated in the figure with dashed vertical lines). On the other hand, due to the attraction between adjacent polymers the side chains slightly stretch or shrink in the RaP3HT, and the geometrical center shifts away from the middle carbon atom. This overall asymmetry in the side chains of RaP3HT also leads to a non-uniform distribution of the local spin states. Hence, the uniformity of the material is distorted which induces an asymmetrically distribution of spin states and the side chains of the RaP3HT rotate asymmetrically along backbone. Similar effects of geometrical asymmetry on the local magnetic states have also been observed previously in atomic C and Si chains.\cite{fan2009density, cahangirov2010long, ozccelik2012self}

\begin{figure}
\includegraphics[width=8cm]{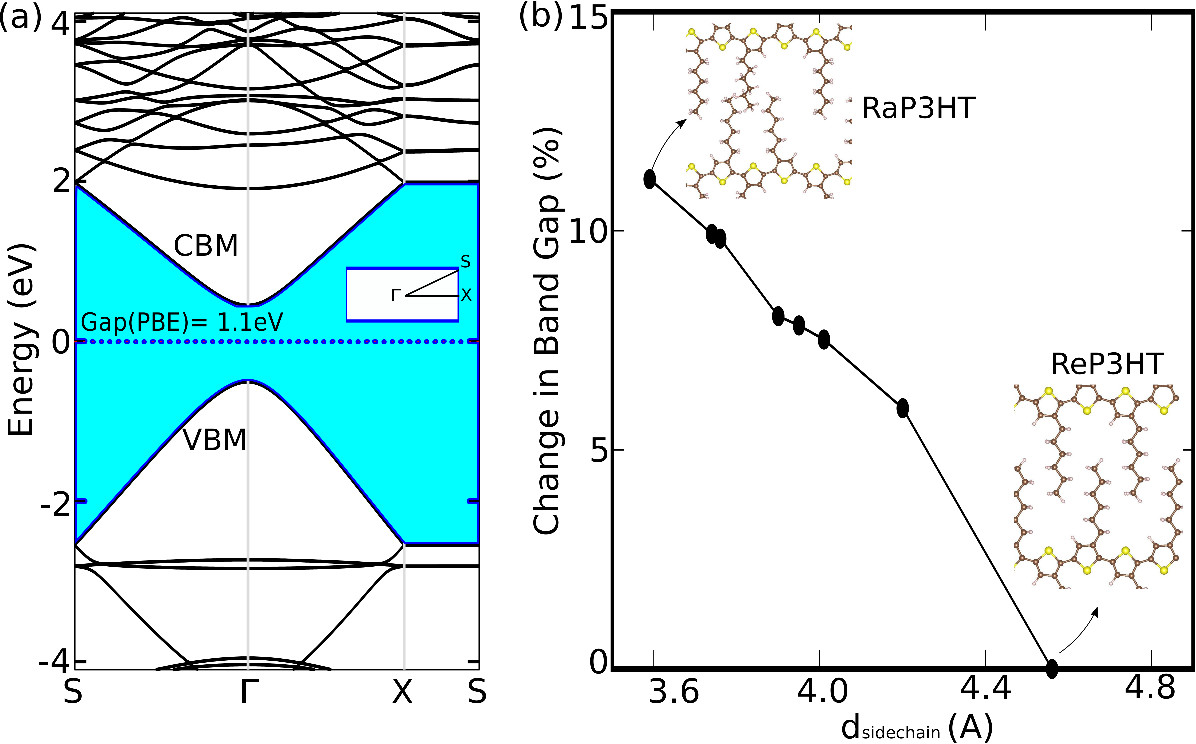}
\caption{(a) Electronic band structure of planar ReP3HT calculated with generalized gradient approximation (GGA). The inset shows the Brillouin Zone and the high symmetry points along which the band structure is calculated. The fundamental bandgap at the $\Gamma$ point increases to 1.6 eV after HSE calculation.  (b) Variation of band gap at the $\Gamma$ point as a function of the shortest distacne between the side chains. }
\label{fig2}
\end{figure}
\begin{figure*}

\includegraphics[width=15cm]{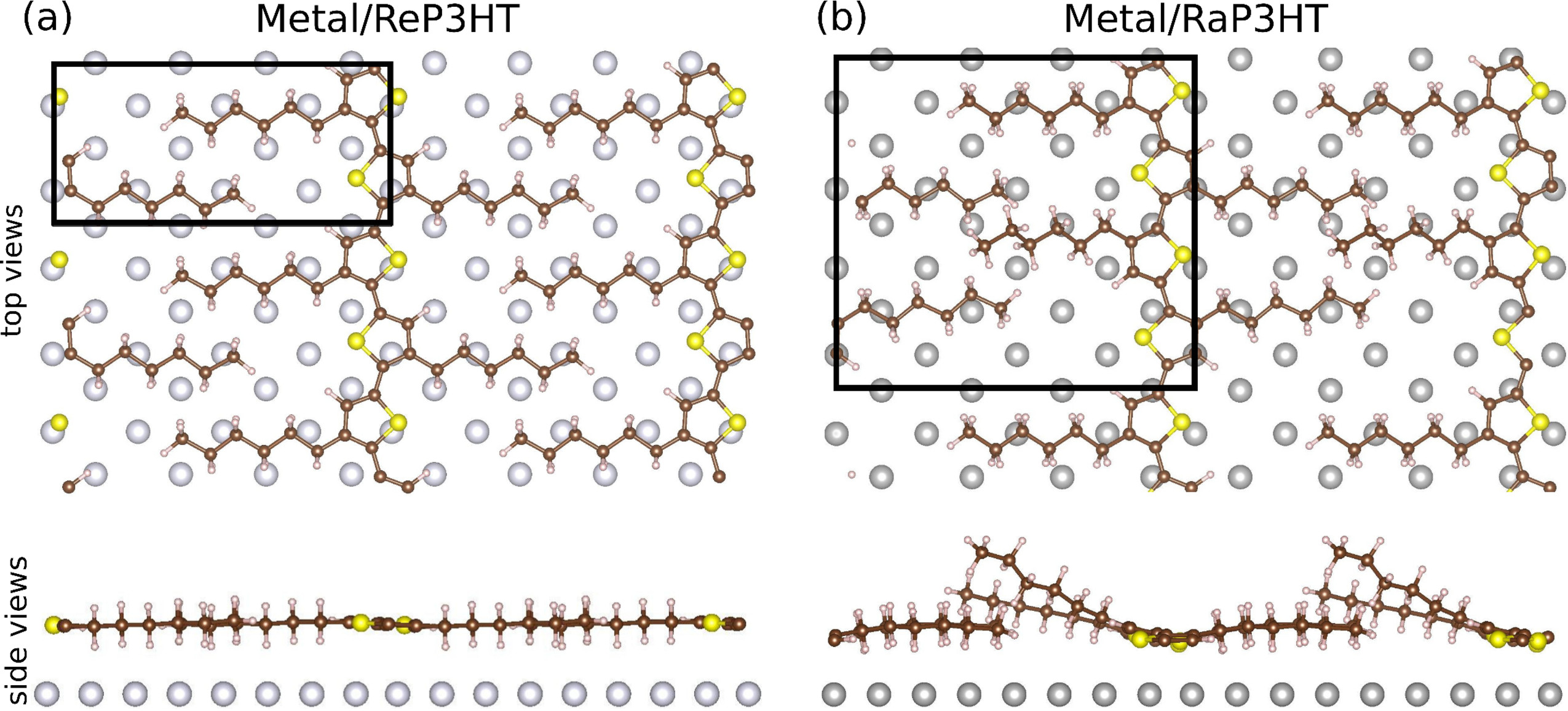}
\caption{(a) Top and side views of the optimized geometries of (a) ReP3HT and (b) RaP3HT on metal substrate where only the atoms in the topmost metal layer is shown. The unit cells are indicated rectangles. The optimized in-plane lattice constants are 7.7 and 16.0 \AA~ for ReP3HT and 15.7 and 16.9 \AA~ for RaP3HT. The side views show the rotation of the sidechains in the metal/RaP3HT system.}
\label{fig3}
\end{figure*}

Rotations of  RaP3HT's side chains bring the heads of adjacent chains closer to each other creating a self-organized but randomly disordered clusters at the contact points of the side chains. Since the charge transfer process in a given material is limited by the most difficult electron hopping process, which is often dominated by disordered regions,\cite{sirringhaus1999two,braun2009energy} the induced regio-randomness has important consequences on the electronic properties of P3HT. In \ref{fig2} we present the electronic structure of ReP3HT and the variation of band gap with increasing regio-randomness which we quantify by measuring the minimum distance between side chains. Planar ReP3HT has a direct band gap of 1.1eV at the $\Gamma$ point where the conduction band minimum (CBM) and valance band maximum (VBM) are mainly composed of the $\pi_z$ states of the backbone. Upon performing Heyd-Scuseria-Ernzerhof (HSE) calculation on top of the GGA, the band gap at the $\Gamma$ point increases to 1.6eV. The electronic energy near the $\Gamma$-point of the BZ changes linearly with respect to \textbf{q}=$\Gamma$-\textbf{k}, which leads to
$ E = \hbar v_F \textbf{q} + O[(\textbf{q} / \textbf{k}^2] $ where $v_F$ is the Fermi velocity. Therefore, by calculating the first derivative of the electronic energies of the $\pi$ bands near the $\Gamma$-point with recpect to \textbf{q}, the Fermi velocity is calculated as $6.94 \times 10^4 m/s$. Similarly, using the curvature of CBM around the $\Gamma$-point, the effective mass defined by $\hbar/m^*$ = $1 / (d^2E/dk^2)$, is calculated as 0.15$m_e$ where $m_e$ is the mass of an electon. Therefore, it is expected that there will be a high-mobility transport within the P3HT structure. As we start to introduce regio-randomness into the system, the band gap gradually increases as shown in \ref{fig2}(b) where the band gap at the $\Gamma$-point is plotted as a function of the minimum distance between the side chains. Although the overall topology of the electronic band structure is conserved, the change in the band gap between the ReP3HT and the structure with highest level of regio-randomness is 11\%.

We next place the Re- and RaP3HT layers on metallic Au, Ag or Pt  crystal surfaces and re-optimize the geometrical and electronic structures to monitor the effect of regio-randomness on the charge transfer at the interface. For RaP3HT, we use the structure with highest band gap. To reduce the computational cost in the current study, we model the metal/P3HT interface only using one layer of P3HT where other P3HT layers can be grown on top of the first layer. In fact, previous studies have reported that the layer-layer distance of multilayer P3HT is large (3.8\AA~ - 4.0\AA~) and  metal - polymer interaction is dominated by the P3HT layer that is nearest to the metallic substrate.\cite{sirringhaus2000microstructure, brinkmann2006orientation} To mimic the semi-infinite metal - polymer interaction, we use three layers of metallic sheets and optimize the interlayer distance between the substrate and P3HT. The optimized structures with Re- and RaP3HT polymers are shown in \ref{fig3}. Note that the side chains of ReP3HT (RaP3HT) maintain their planar (rotated) orientations on the metallic substrate where the backbones are  planar in both cases. The surface binding energy per unitcell is calculated from the expression $E_B=E_{P3HT} + E_{metal} - E_{[P3HT+metal]}$, in terms of the optimized total energies of isolated P3HT, substrate and the combined metal/P3HT heterostructure where a positive value indicates adhesive binding.  Accordingly, the binding energies between the substrates and the polymers per thiophene ring  in the unitcells presented in \ref{fig3}(a-b) were calculated as 2.92, 2.88, 2.11 eV for ReP3HT and 1.45, 1.43, 1.01 eV for RaP3HT on Ag, Au and Pt surfaces, respectively.  These values indicate that P3HT has a strong  stability on the metal surfaces and the magnitude of the binding energies are comparable with those of similar structures which were reported in the literature  as 1.42 eV for P3HT on ZnO surface and 2.17 eV for a only a single S-Au bond.\cite{li2014atomic, pensa2012chemistry} To investigate the effects of side chains on the strength of interfacial binding, we removed the side chains from the ReP3HT and recalculated its binding energy on Au surface which dropped from 2.88 eV to 1.53 eV. This shows that both the side chains and the backbone  contribute significantly to the adhesion of P3HT polymer on the metal substrate, as also reported previously for the NiO/P3HT interface.\cite{li2014atomic} We note that, althoguh the side chains do not perturb the backbone, they strengthen the surface binding energy between the metallic substrate and P3HT due to the increase in the contact area at the interface. This also explains the relatively lower binding energy of the RaP3HT polymer as compared to ReP3HT, since the side chains of P3HT rotate away from the substrate in the metal/RaP3HT system which decreases interfacial contact area. 

\begin{figure}
\includegraphics[width=8cm]{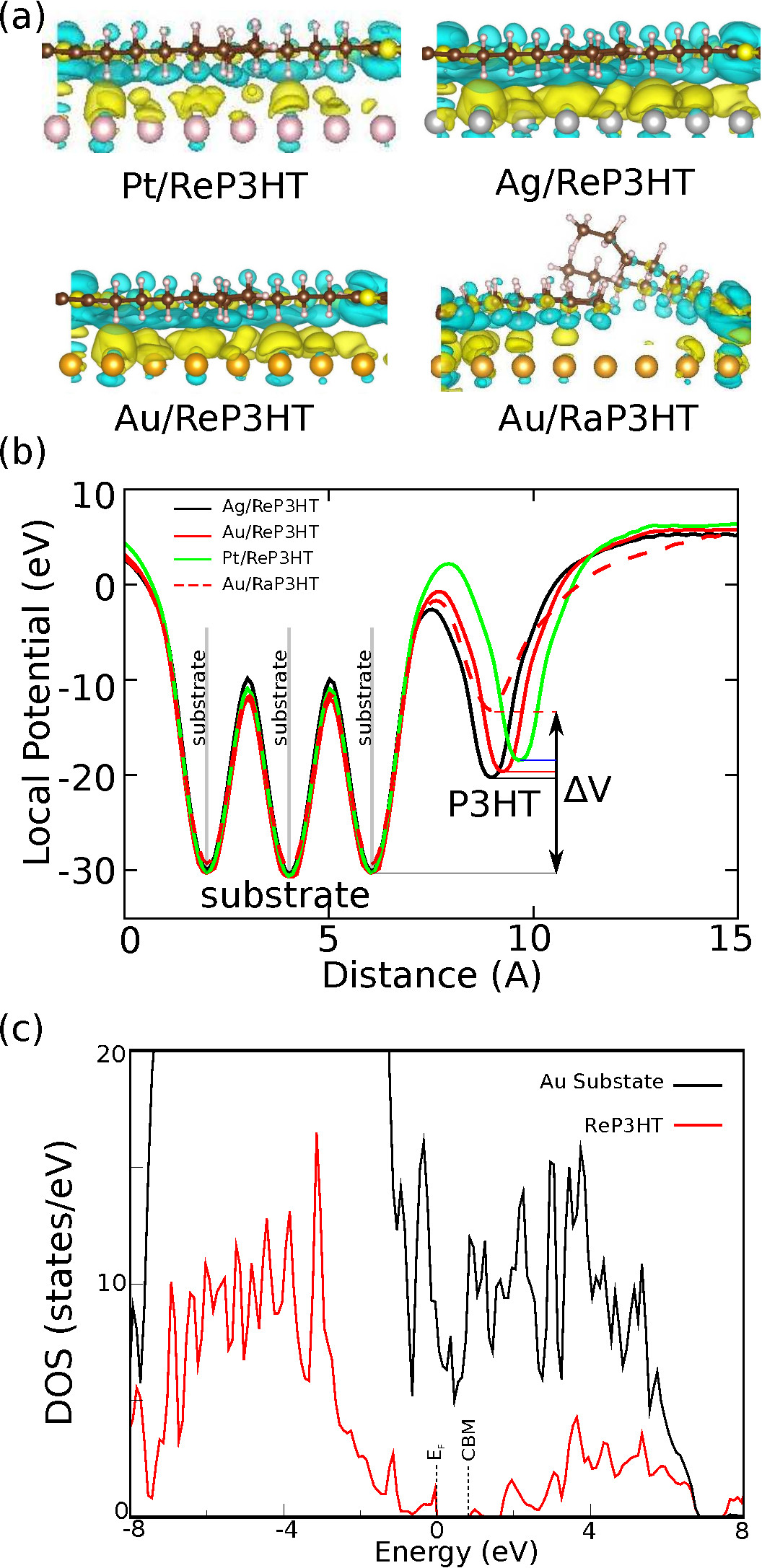}
\caption{(a) Charge transfer isosurfaces for various metal/P3HT systems. Only the top layer of the substrate is shown. (b) Variation of the electrostatic local potential along the out-of-plane direction of the metal/P3HT systems shown in (a). (c) Density of states projected onto P3HT and Au atoms in the Au/ReP3HT system.}
\label{fig4}
\end{figure}

To assess the level of hybridization in these metal/P3HT heterostructures, we calculate  electronic wavefunctions and  density of states (DOS) at their interfaces. The charge distribution upon the formation each metal/P3HT interface was obtained by calculating the difference of the local charge densities of the heterostructure and individual materials using  $\Delta \rho = \rho_{[metal/P3HT]} - \rho_{[P3HT]} - \rho_{[metal]} $ where each term of this equation is calculated separately in the same unitcell. The volumetric charge difference plots shown in \ref{fig4}(a) were calculated using the same isosurface value for the sake of comparability. In each case, the substantial amount of electron-hole accumulation observed in the interlayer region indicates strong hybridization. Charge density populations per area on the surfaces were calculated as 0.061, 0.056, and 0.038 C/m$^2$ for Ag/ReP3HT, Au/ReP3HT and Pt/ReP3HT interfaces, respectively. These numbers drop to 0.051, 0.040, and 0.027 C/m$^2$ for Ag/RaP3HT, Au/RaP3HT and Pt/RaP3HT interfaces. The order of the surface charge density values in Ag, Au and Pt substrates is similar to the trend in the binding energies of P3HT on these substrates such that higher charge transfer leads to stronger binding. The decrease in the amount of charge transfer in the interface after inducing regio-randomness to the side chains of the P3HT polymer can be seen from the difference between Au/ReP3HT and Au/RaP3HT charge transfer plots in \ref{fig4}(a).

The level of charge transfer at the interfaces also manifests itself in the distribution of the DFT computed local electrostatic potential difference as shown in \ref{fig4}(b). The difference between the potential of the metallic surface and ReP3HT is lower for the heterostructures with higher amount of charge transfer (namely, $\Delta V_{Ag-Re3HT} < \Delta V_{Au-ReP3HT} < \Delta V_{Pt-ReP3HT}$ where replacing  ReP3HT with RaP3HT also significantly increases the potential difference as shown for the Au substrate. Using the local electrostatic potential plots, the work functions of the interfaces were calculated by $\Phi_{workfunction} = E_{vacuum} - E_{Fermi}$ where $E_{vacuum}$ is vacuum potential and $E_{Fermi}$ is the energy of the highest occupied state of each system obtained from electronic structure calculations. Accordingly, the work functions of ReP3HT on Ag, Au, and Pt substrates are 3.43 eV, 4.19 eV and 5.45 eV, respectively. Here again, the interface with the most charge transfer has the lowest work function since charge transfer is lowering the vacuum level towards the Fermi level, and the amount of work needed to remove an electron from the surface decreased. The work function of the isolated gold substrate was calculated as 5.55 eV which indicates that the work function of the metal substrate can be engineered by depositing organic films to introduce charge transfers. The densities of electronic states are also effected by the charge transfer process at the interface as shown in \ref{fig4}(c) for Au/ReP3HT.  Accordingly, Au states fill the gap region of the isolated P3HT layer and the Fermi level of the system is just above the HOMO state of P3HT, which is set to zero in \ref{fig4}(c). In the heterostructure, the first GGA calculated CBM state of ReP3HT appears at 0.7 eV (which is 0.4 eV lower than the GGA calculated CBM energy of the freestanding ReP3HT). Thus, electron transfer from the P3HT layer to  Au states is easily possible and it leads to a strong hybridization at the interface of the heterostructure. This is a direct indication of strong electronic coupling at the interface and high charge generation efficiency for photovoltaic applications where electrons can be pumped towards the conduction bands upon exciton diffusion to the interface. We note that, although GGA underestimates the band gap, we see that the band gap of P3HT decreases by 0.4 eV after it is deposited on the gold substrate. This is consistent with our experimental results, where we found that the initial band gap of P3HT  is 2.1 eV and by depositing P3HT on Au, we create an interfacial gap of 1.5 eV.\cite{xiang2017ultrafast}

\begin{figure}
\includegraphics[width=8cm]{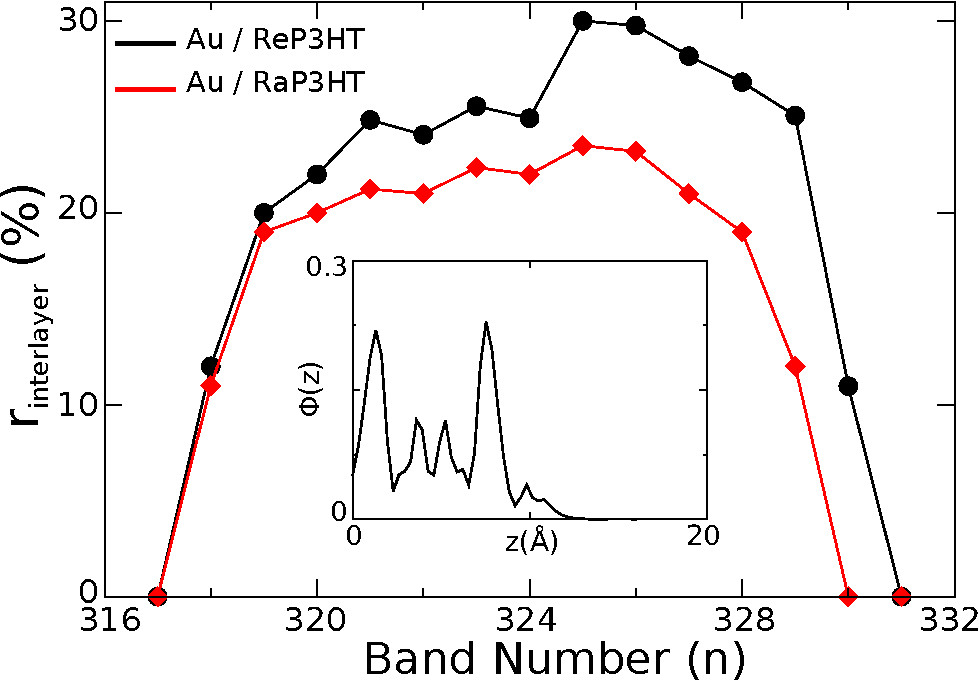}
\caption{Percentage of the interlayer states localized in the interlayer region of the Au/P3HT and Au/RaP3HT heterostructures on different bands are shown with black and red curves, respectively. The inset shows the probability density function, $\Phi(z)$, at the conduction band minimum of Au/ReP3HT.}
\label{fig5}
\end{figure}

We further evaluate the level of hybridization by calculating the real space wavefunction along the material. Using the DFT calculated wave function data which includes the lattice points in the reciprocal space and their corresponding plane wave coefficients, real-space wavefunctions were constructed. Starting from  the pseudo-wave function, $\Psi_{n,k}(r)$, of a specific band (n) at a specified k-point, the probability function was extracted to a specified region in the real space as $\Phi = \int_{z_1}^{z_2} dz \int_{\Omega} \Psi_{n,k}(r)^2 dx dy$, where $\Omega$ is the two-dimensional cross section of the unit cell, and $z_1$-$z_2$ are the boundaries in the out-of-plane direction where the wavefunction is evaluated.\cite{feenstra2013low, ozccelik2018tin, chaves2018electrical} Here, we focused on the wave function at the $\Gamma$ point since that is where CBM and VBM of P3HT reside. We performed these calculations for different bands at the $\Gamma$ point and calculate the percentage of states that are localized in the interlayer region as shown in \ref{fig5}, where r$_{interlayer}$ is calculated by adding the probability functions in the interlayer region. Accordingly, ratio of states residing in the interface is higher in the Au/ReP3HT system as compared to Au/RaP3Ht where both heterostructures exhibit strong hybridization and interlayer coupling of electronic states.

\begin{figure*}
\includegraphics[width=15cm]{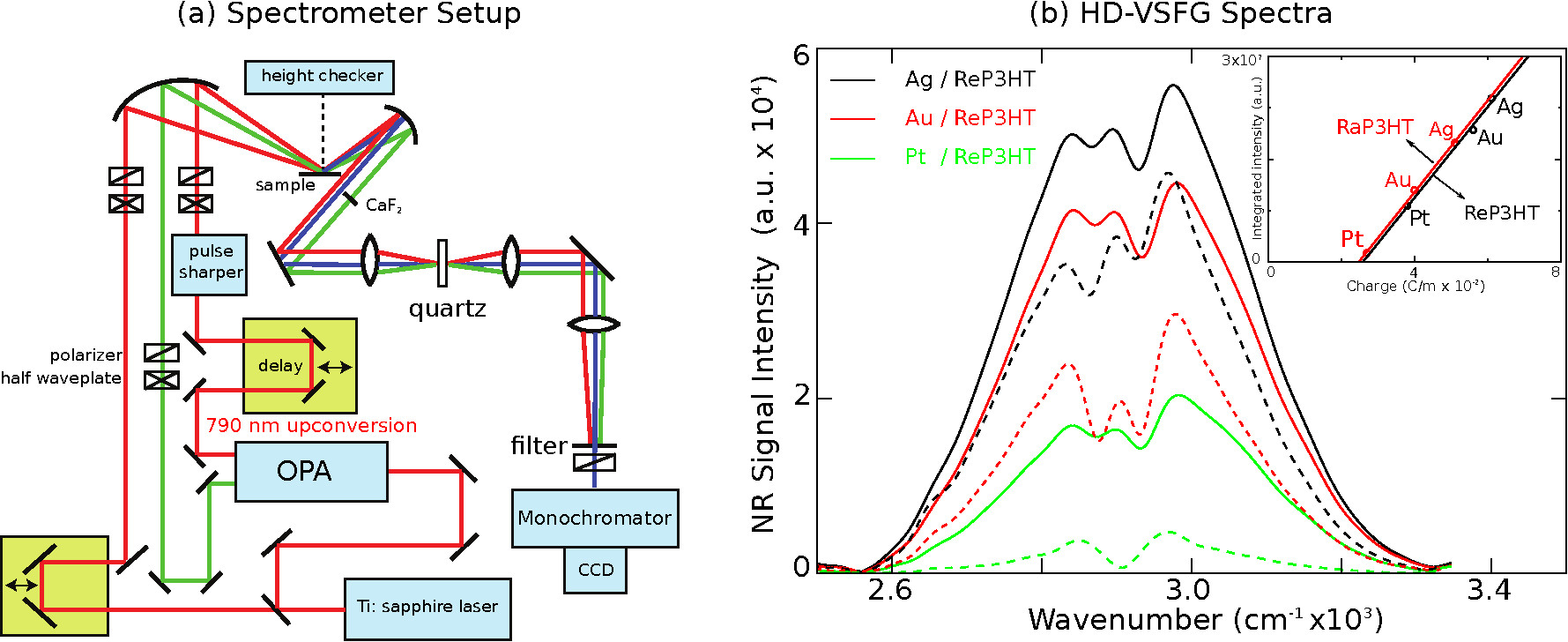}
\caption{(a) Schematic description of the spectrometer that is used to collect the HD-VSFG spectra. (b) The nonresonant part of the HD-VSFG spectra of ReP3HT and RaP3HT polymers on different substrates are shown with solid and dashed curves, respectively. The inset shows the linear variation of the integrated nonresonant signal intensity with charge density at the interface.}
\label{fig6}
\end{figure*}

We experimentally investigate the effect of disorder on metal/P3HT interfaces using HD-VSFG which has been shown to be a sensitive tool for probing molecular structures at interfaces.\cite{zhuang1999mapping, hore2005whole, o2015experimental, wang2015quantitative, li2015probing} Here, we collect HD-VSFG spectra of different metal/P3HT systems using the spectrometer\cite{xiang2017ultrafast, li2019heterodyne} that is schematically presented in \ref{fig6}(a) and the results  are shown in \ref{fig6}(b). Accordingly, ReP3HT on all three metal substrates show significant amount of nonresonant signal where the Ag/P3HT system has the largest nonresonant signal followed by Au/P3HT and Pt/P3HT. We note that  since all of the  HD-VSFG data were taken under SSP polarization condition, there is no nonresonant signal by metal substrates themselves due to the large lateral screening of free electron motions. The fact that nonresonant signal becomes strong at metal/P3HT interfaces, indicates that there is a substantiate modification of the interfacial electronic structure.  Furthermore, this nonresonant signal intensity trend agrees with the spontaneous charge transfer calculation shown in \ref{fig4}(a) which suggests the origin of nonresonant signal is related to the charge redistribution across the interface. On the other hand, metal/RaP3HT systems have an overall smaller nonresonant signal, especially for Pt/RaP3HT where there is almost no nonresonant signal which further verifies the correlation between the spectral features to the electronic structure. The intensity of nonresonant signal show a linear correlation with the calculated spontaneous interfacial charge transfer as shown in the inset of \ref{fig6}(b). There are two possible origins of the nonresonant signal in HD-VSFG spectra. First, the hybridization between the organic polymer and metal electronic orbitals significantly reduce the lateral screening force, e.g. the dielectric constant of P3HT is significantly smaller than metal. Therefore, lateral free electron oscillation driven by the external electromagnetic field are not completely screened and SSP SFG signal becomes allowed. Second, interfacial charge transfer generates an interfacial electric field, which subsequently lead to a $\chi^{(3)}$ contribution of the nonresonant signal.\cite{ohno2017second} Further investigation will  be necessary to completely explain the mechanism of spontaneous interfacial charge transfer induced nonresonant signal.

In conclusion, by combining ab-initio quantum chemical calculations based on DFT with HD-VSFG measurements, we showed that the geometrical conformation of the organic polymer at a metal/organic heterojunction is significant at determining the electronic properties of the system and the amount of charge transfer between the layers. We show that the electronic coupling at the interface and the work function of the material can be selectively engineered by modifying the level of regio-randomness present in the side chains of the P3HT polymer where increasing regio-randomness lowers the charge transfer and electronic coupling. We revealed that this effect is a direct consequence of the modification of the real space wave function of the material and hybridization of its electronic states. The theoretical predictions presented here are in agreement with our HD-VSFG measurements where both of which suggest that there is a direct correlation with the level of charge transfer and the intensity of the nonresonant response of the metal/P3HT systems.  Thus, by measuring the non-resonant response of the metal/P3HT heterojunctions, we show that the level of spontaneous charge transfer at the interface can be determined. Our results pave the way of a new theoretical and experimental approach for probing the spontaneous charge transfer under ambient condition.

\paragraph*{Methods:}
First principles calculations  were performed using the generalized gradient approximation including van der Waals corrections. \cite{grimme2006semi} Projector-augmented wave potentials\cite{blochl94} were used and the exchange-correlation potential was approximated with the Perdew-Burke-Ernzerhof(PBE) functional.\cite{pbe} The Brillouin zone(BZ) was sampled in the Monkhorst-Pack scheme where the \textbf{k}-point sampling of (11$\times$11$\times$1) was found to be suitable for the BZ corresponding to the primitive unit cell of P3HT. For larger unit cells, the k-point sampling was scaled accordingly. While modelling the heterostructures, the size of the periodic unit cell in the out-of-plane direction was tested to be sufficiently large to avoid coupling between periodic units and a vacuum spacing of 10 \AA~ was used. A plane-wave basis set with energy cutoff value of 450 eV was used where the energy convergence value between two consecutive steps was chosen as $10^{-5}$ eV. A maximum force of 0.1 eV/\AA~ was allowed on each atom. The electronic band structures were calculated using the pre-converged unit cells and charge distributions. The bands were calculated along the $S - \Gamma - X - S$ high symmetry points in the first BZ where 50 \textbf{k}-points were used between each high symmetry point. Numerical calculations were carried out using the VASP software.\cite{vasp}

The metal/P3HT interfaces used in HD-VSFG experiments were prepared by spin coating method at 1500 rpm. Regioregular (REIKE Metal) and regiorandom P3HT (Sigma Aldrich) were dissolved in chloroform with the concentration of 8 mg/ml respectively. The metal substrates were prepared by magnetron sputtering where a titanium layer was firstly sputtered onto glass slice to improve the adhesion between metals and glass substrates, followed by depositing gold, silver, and platinum on top where the thickness is determined to be 150 nm.

For the HD-VSFG experiments, the light source is an ultrafast Ti:sapphire regenerative amplifier (Astrella, Coherent) that outputs 790 nm pulses with 5W power. The 790 nm beam is then sent into an optical parametric amplifier (TOPAS, Light Conversion) to generate two tunable near-IR pulses which are referred as signal and idler beam. The two near-IR beams are then spatially and temporally overlapped on a type I BBO crystal to generate the mid-IR beam that centers at 3.3 μm to be on resonant with the C-H vibrational mode in P3HT. The residual 790 nm beam after TOPAS is used as upconversion beam which passes through a spectral pulse shaper to narrow down the spectral bandwidth to 0.6 nm or 9.5 cm-1 at full-width half max (FWHM) to improve the spectral resolution. The HD-VSFG signal is generated by spatially and temporally overlapping the mid-IR pulse and upconversion pulse at the sample surface. The reflected mid-IR pulse then passes through a 2-mm thick CaF2 delay media to generate a time delay relative to the HD-VSFG signal. The upconversion beam and delayed mid-IR beam are then focused onto a Y-cut quartz crystal by a lens (f=10 cm) to generate the local oscillator (LO) for heterodyne detection. Both LO and HD-VSFG signals are collimated by another lens (f=10 cm) and then transmit through a short-band pass filter to remove the 790 nm residual. After passing through the monochromator, the LO and HD-VSFG signals interfere with each other at a charge-coupled device (CCD) to generate the heterodyne HD-VSFG signal. 

\paragraph*{Acknowledgment:}

This research was supported by the National Science Foundation (NSF) through grant CHE-1808111. All computations were performed using the resources of the Extreme Science and Engineering Discovery Environment (XSEDE), which is supported by the NSF through grant ACI-1053575, under allocation TG-CHE110009.

\bibliography{arxiv.bbl}

\providecommand{\latin}[1]{#1}
\providecommand*\mcitethebibliography{\thebibliography}
\csname @ifundefined\endcsname{endmcitethebibliography}
  {\let\endmcitethebibliography\endthebibliography}{}
\begin{mcitethebibliography}{37}
\providecommand*\natexlab[1]{#1}
\providecommand*\mciteSetBstSublistMode[1]{}
\providecommand*\mciteSetBstMaxWidthForm[2]{}
\providecommand*\mciteBstWouldAddEndPuncttrue
  {\def\EndOfBibitem{\unskip.}}
\providecommand*\mciteBstWouldAddEndPunctfalse
  {\let\EndOfBibitem\relax}
\providecommand*\mciteSetBstMidEndSepPunct[3]{}
\providecommand*\mciteSetBstSublistLabelBeginEnd[3]{}
\providecommand*\EndOfBibitem{}
\mciteSetBstSublistMode{f}
\mciteSetBstMaxWidthForm{subitem}{(\alph{mcitesubitemcount})}
\mciteSetBstSublistLabelBeginEnd
  {\mcitemaxwidthsubitemform\space}
  {\relax}
  {\relax}

\bibitem[Bae \latin{et~al.}(2019)Bae, Kum, Kong, Kim, Choi, Lee, Lin, Park, and
  Kim]{bae2019integration}
Bae,~S.-H.; Kum,~H.; Kong,~W.; Kim,~Y.; Choi,~C.; Lee,~B.; Lin,~P.; Park,~Y.;
  Kim,~J. Integration of bulk materials with two-dimensional materials for
  physical coupling and applications. \emph{Nat. Mater.} \textbf{2019},
  \emph{18}, 550\relax
\mciteBstWouldAddEndPuncttrue
\mciteSetBstMidEndSepPunct{\mcitedefaultmidpunct}
{\mcitedefaultendpunct}{\mcitedefaultseppunct}\relax
\EndOfBibitem
\bibitem[Huynh \latin{et~al.}(2002)Huynh, Dittmer, and
  Alivisatos]{huynh2002hybrid}
Huynh,~W.~U.; Dittmer,~J.~J.; Alivisatos,~A.~P. Hybrid nanorod-polymer solar
  cells. \emph{Science} \textbf{2002}, \emph{295}, 2425--2427\relax
\mciteBstWouldAddEndPuncttrue
\mciteSetBstMidEndSepPunct{\mcitedefaultmidpunct}
{\mcitedefaultendpunct}{\mcitedefaultseppunct}\relax
\EndOfBibitem
\bibitem[Yin and Alivisatos(2004)Yin, and Alivisatos]{yin2004colloidal}
Yin,~Y.; Alivisatos,~A.~P. Colloidal nanocrystal synthesis and the
  organic--inorganic interface. \emph{Nature} \textbf{2004}, \emph{437},
  664\relax
\mciteBstWouldAddEndPuncttrue
\mciteSetBstMidEndSepPunct{\mcitedefaultmidpunct}
{\mcitedefaultendpunct}{\mcitedefaultseppunct}\relax
\EndOfBibitem
\bibitem[Coakley and McGehee(2004)Coakley, and McGehee]{coakley2004conjugated}
Coakley,~K.~M.; McGehee,~M.~D. Conjugated polymer photovoltaic cells.
  \emph{Chem. Mater.} \textbf{2004}, \emph{16}, 4533--4542\relax
\mciteBstWouldAddEndPuncttrue
\mciteSetBstMidEndSepPunct{\mcitedefaultmidpunct}
{\mcitedefaultendpunct}{\mcitedefaultseppunct}\relax
\EndOfBibitem
\bibitem[Chang \latin{et~al.}(2010)Chang, Rhee, Im, Lee, Kim, Seok,
  Nazeeruddin, and Gratzel]{chang2010high}
Chang,~J.~A.; Rhee,~J.~H.; Im,~S.~H.; Lee,~Y.~H.; Kim,~H.-j.; Seok,~S.~I.;
  Nazeeruddin,~M.~K.; Gratzel,~M. High-performance nanostructured inorganic-
  organic heterojunction solar cells. \emph{Nano Lett.} \textbf{2010},
  \emph{10}, 2609--2612\relax
\mciteBstWouldAddEndPuncttrue
\mciteSetBstMidEndSepPunct{\mcitedefaultmidpunct}
{\mcitedefaultendpunct}{\mcitedefaultseppunct}\relax
\EndOfBibitem
\bibitem[Braun \latin{et~al.}(2009)Braun, Salaneck, and
  Fahlman]{braun2009energy}
Braun,~S.; Salaneck,~W.~R.; Fahlman,~M. Energy-level alignment at organic/metal
  and organic/organic interfaces. \emph{Adv. Mater.} \textbf{2009}, \emph{21},
  1450--1472\relax
\mciteBstWouldAddEndPuncttrue
\mciteSetBstMidEndSepPunct{\mcitedefaultmidpunct}
{\mcitedefaultendpunct}{\mcitedefaultseppunct}\relax
\EndOfBibitem
\bibitem[Dag and Wang(2008)Dag, and Wang]{dag2008modeling}
Dag,~S.; Wang,~L.-W. Modeling of Nanoscale morphology of regioregular poly
  (3-hexylthiophene) on a ZnO (1010) surface. \emph{Nano Lett.} \textbf{2008},
  \emph{8}, 4185--4190\relax
\mciteBstWouldAddEndPuncttrue
\mciteSetBstMidEndSepPunct{\mcitedefaultmidpunct}
{\mcitedefaultendpunct}{\mcitedefaultseppunct}\relax
\EndOfBibitem
\bibitem[Dag and Wang(2010)Dag, and Wang]{dag2010packing}
Dag,~S.; Wang,~L.-W. Packing structure of poly (3-hexylthiophene) crystal: ab
  initio and molecular dynamics studies. \emph{J. Phys. Chem. C} \textbf{2010},
  \emph{114}, 5997--6000\relax
\mciteBstWouldAddEndPuncttrue
\mciteSetBstMidEndSepPunct{\mcitedefaultmidpunct}
{\mcitedefaultendpunct}{\mcitedefaultseppunct}\relax
\EndOfBibitem
\bibitem[Dang \latin{et~al.}(2011)Dang, Hirsch, and Wantz]{dang2011p3ht}
Dang,~M.~T.; Hirsch,~L.; Wantz,~G. P3HT: PCBM, best seller in polymer
  photovoltaic research. \emph{Adv. Mater.} \textbf{2011}, \emph{23},
  3597--3602\relax
\mciteBstWouldAddEndPuncttrue
\mciteSetBstMidEndSepPunct{\mcitedefaultmidpunct}
{\mcitedefaultendpunct}{\mcitedefaultseppunct}\relax
\EndOfBibitem
\bibitem[Holliday \latin{et~al.}(2016)Holliday, Ashraf, Wadsworth, Baran,
  Yousaf, Nielsen, Tan, Dimitrov, Shang, and et. al.]{holliday2016high}
Holliday,~S.; Ashraf,~R.~S.; Wadsworth,~A.; Baran,~D.; Yousaf,~S.~A.;
  Nielsen,~C.~B.; Tan,~C.-H.; Dimitrov,~S.~D.; Shang,~Z.; et. al.,
  High-efficiency and air-stable P3HT-based polymer solar cells with a new
  non-fullerene acceptor. \emph{Nat. Commun.} \textbf{2016}, \emph{7},
  11585\relax
\mciteBstWouldAddEndPuncttrue
\mciteSetBstMidEndSepPunct{\mcitedefaultmidpunct}
{\mcitedefaultendpunct}{\mcitedefaultseppunct}\relax
\EndOfBibitem
\bibitem[Wang \latin{et~al.}(2004)Wang, Moses, Heeger, Zhang, Narasimhan, and
  Demaray]{wang2004poly}
Wang,~G.; Moses,~D.; Heeger,~A.~J.; Zhang,~H.-M.; Narasimhan,~M.; Demaray,~R.
  Poly (3-hexylthiophene) field-effect transistors with high dielectric
  constant gate insulator. \emph{J. Appl. Phys.} \textbf{2004}, \emph{95},
  316--322\relax
\mciteBstWouldAddEndPuncttrue
\mciteSetBstMidEndSepPunct{\mcitedefaultmidpunct}
{\mcitedefaultendpunct}{\mcitedefaultseppunct}\relax
\EndOfBibitem
\bibitem[Sirringhaus \latin{et~al.}(1999)Sirringhaus, Brown, Friend, Nielsen,
  Bechgaard, Langeveld-Voss, Spiering, Janssen, Meijer, Herwig, and
  Leeuw]{sirringhaus1999two}
Sirringhaus,~H.; Brown,~P.; Friend,~R.; Nielsen,~M.~M.; Bechgaard,~K.;
  Langeveld-Voss,~B.; Spiering,~A.; Janssen,~R.~A.; Meijer,~E.; Herwig,~P.;
  Leeuw,~D.~M. Two-dimensional charge transport in self-organized,
  high-mobility conjugated polymers. \emph{Nature} \textbf{1999}, \emph{401},
  685\relax
\mciteBstWouldAddEndPuncttrue
\mciteSetBstMidEndSepPunct{\mcitedefaultmidpunct}
{\mcitedefaultendpunct}{\mcitedefaultseppunct}\relax
\EndOfBibitem
\bibitem[Anglin \latin{et~al.}(2010)Anglin, O'Brien, and
  Massari]{anglin2010monitoring}
Anglin,~T.~C.; O'Brien,~D.~B.; Massari,~A.~M. Monitoring the charge
  accumulation process in polymeric field-effect transistors via in situ sum
  frequency generation. \emph{J. Phys. Chem. C} \textbf{2010}, \emph{114},
  17629--17637\relax
\mciteBstWouldAddEndPuncttrue
\mciteSetBstMidEndSepPunct{\mcitedefaultmidpunct}
{\mcitedefaultendpunct}{\mcitedefaultseppunct}\relax
\EndOfBibitem
\bibitem[Xiang \latin{et~al.}(2017)Xiang, Li, Pham, Paesani, and
  Xiong]{xiang2017ultrafast}
Xiang,~B.; Li,~Y.; Pham,~C.~H.; Paesani,~F.; Xiong,~W. Ultrafast direct
  electron transfer at organic semiconductor and metal interfaces. \emph{Sci.
  Adv.} \textbf{2017}, \emph{3}, e1701508\relax
\mciteBstWouldAddEndPuncttrue
\mciteSetBstMidEndSepPunct{\mcitedefaultmidpunct}
{\mcitedefaultendpunct}{\mcitedefaultseppunct}\relax
\EndOfBibitem
\bibitem[Lindau and Spicer(1974)Lindau, and Spicer]{lindau1974probing}
Lindau,~I.; Spicer,~W. The probing depth in photoemission and Auger-electron
  spectroscopy. \emph{J. Electron. Spectrosc. Relat. Phenom} \textbf{1974},
  \emph{3}, 409--413\relax
\mciteBstWouldAddEndPuncttrue
\mciteSetBstMidEndSepPunct{\mcitedefaultmidpunct}
{\mcitedefaultendpunct}{\mcitedefaultseppunct}\relax
\EndOfBibitem
\bibitem[Sirringhaus \latin{et~al.}(2000)Sirringhaus, Brown, Friend, Nielsen,
  Bechgaard, Langeveld-Voss, Spiering, Janssen, and
  Meijer]{sirringhaus2000microstructure}
Sirringhaus,~H.; Brown,~P.; Friend,~R.; Nielsen,~M.~M.; Bechgaard,~K.;
  Langeveld-Voss,~B.; Spiering,~A.; Janssen,~R.; Meijer,~E.
  Microstructure--mobility correlation in self-organised, conjugated polymer
  field-effect transistors. \emph{Synth. Met.} \textbf{2000}, \emph{111},
  129--132\relax
\mciteBstWouldAddEndPuncttrue
\mciteSetBstMidEndSepPunct{\mcitedefaultmidpunct}
{\mcitedefaultendpunct}{\mcitedefaultseppunct}\relax
\EndOfBibitem
\bibitem[Fan \latin{et~al.}(2009)Fan, Liu, Lin, Shen, and Kuo]{fan2009density}
Fan,~X.; Liu,~L.; Lin,~J.; Shen,~Z.; Kuo,~J.-L. Density functional theory study
  of finite carbon chains. \emph{ACS Nano} \textbf{2009}, \emph{3},
  3788--3794\relax
\mciteBstWouldAddEndPuncttrue
\mciteSetBstMidEndSepPunct{\mcitedefaultmidpunct}
{\mcitedefaultendpunct}{\mcitedefaultseppunct}\relax
\EndOfBibitem
\bibitem[Cahangirov \latin{et~al.}(2010)Cahangirov, Topsakal, and
  Ciraci]{cahangirov2010long}
Cahangirov,~S.; Topsakal,~M.; Ciraci,~S. Long-range interactions in carbon
  atomic chains. \emph{Phys. Rev. B} \textbf{2010}, \emph{82}, 195444\relax
\mciteBstWouldAddEndPuncttrue
\mciteSetBstMidEndSepPunct{\mcitedefaultmidpunct}
{\mcitedefaultendpunct}{\mcitedefaultseppunct}\relax
\EndOfBibitem
\bibitem[{\"O}z{\c{c}}elik and Ciraci(2012){\"O}z{\c{c}}elik, and
  Ciraci]{ozccelik2012self}
{\"O}z{\c{c}}elik,~V.~O.; Ciraci,~S. Self-assembly mechanisms of short atomic
  chains on single-layer graphene and boron nitride. \emph{Phys. Rev. B}
  \textbf{2012}, \emph{86}, 155421\relax
\mciteBstWouldAddEndPuncttrue
\mciteSetBstMidEndSepPunct{\mcitedefaultmidpunct}
{\mcitedefaultendpunct}{\mcitedefaultseppunct}\relax
\EndOfBibitem
\bibitem[Brinkmann and Wittmann(2006)Brinkmann, and
  Wittmann]{brinkmann2006orientation}
Brinkmann,~M.; Wittmann,~J.-C. Orientation of regioregular poly
  (3-hexylthiophene) by directional solidification: a simple method to reveal
  the semicrystalline structure of a conjugated polymer. \emph{Adv. Mater.}
  \textbf{2006}, \emph{18}, 860--863\relax
\mciteBstWouldAddEndPuncttrue
\mciteSetBstMidEndSepPunct{\mcitedefaultmidpunct}
{\mcitedefaultendpunct}{\mcitedefaultseppunct}\relax
\EndOfBibitem
\bibitem[Li \latin{et~al.}(2014)Li, Kontsevoi, and Freeman]{li2014atomic}
Li,~L.-H.; Kontsevoi,~O.~Y.; Freeman,~A.~J. Atomic-scale understanding of the
  interaction of poly (3-hexylthiophene) with the NiO (100) surface: A
  first-principles study. \emph{J. Phys. Chem. C} \textbf{2014}, \emph{118},
  20298--20305\relax
\mciteBstWouldAddEndPuncttrue
\mciteSetBstMidEndSepPunct{\mcitedefaultmidpunct}
{\mcitedefaultendpunct}{\mcitedefaultseppunct}\relax
\EndOfBibitem
\bibitem[Pensa \latin{et~al.}(2012)Pensa, Cortes, Corthey, Carro, Vericat,
  Fonticelli, Benitez, Rubert, and Salvarezza]{pensa2012chemistry}
Pensa,~E.; Cortes,~E.; Corthey,~G.; Carro,~P.; Vericat,~C.; Fonticelli,~M.~H.;
  Benitez,~G.; Rubert,~A.~A.; Salvarezza,~R.~C. The chemistry of the
  sulfur--gold interface: in search of a unified model. \emph{Accounts Chem.
  Res.} \textbf{2012}, \emph{45}, 1183--1192\relax
\mciteBstWouldAddEndPuncttrue
\mciteSetBstMidEndSepPunct{\mcitedefaultmidpunct}
{\mcitedefaultendpunct}{\mcitedefaultseppunct}\relax
\EndOfBibitem
\bibitem[Feenstra \latin{et~al.}(2013)Feenstra, Srivastava, Gao, Widom,
  Diaconescu, Ohta, Kellogg, Robinson, and Vlassiouk]{feenstra2013low}
Feenstra,~R.~M.; Srivastava,~N.; Gao,~Q.; Widom,~M.; Diaconescu,~B.; Ohta,~T.;
  Kellogg,~G.; Robinson,~J.; Vlassiouk,~I. Low-energy electron reflectivity
  from graphene. \emph{Phys. Rev. B} \textbf{2013}, \emph{87}, 041406\relax
\mciteBstWouldAddEndPuncttrue
\mciteSetBstMidEndSepPunct{\mcitedefaultmidpunct}
{\mcitedefaultendpunct}{\mcitedefaultseppunct}\relax
\EndOfBibitem
\bibitem[{\"O}z{\c{c}}elik \latin{et~al.}(2018){\"O}z{\c{c}}elik, Fathi,
  Azadani, and Low]{ozccelik2018tin}
{\"O}z{\c{c}}elik,~V.~O.; Fathi,~M.; Azadani,~J.~G.; Low,~T. Tin
  monochalcogenide heterostructures as mechanically rigid infrared band gap
  semiconductors. \emph{Phys. Rev. Mater.} \textbf{2018}, \emph{2},
  051003\relax
\mciteBstWouldAddEndPuncttrue
\mciteSetBstMidEndSepPunct{\mcitedefaultmidpunct}
{\mcitedefaultendpunct}{\mcitedefaultseppunct}\relax
\EndOfBibitem
\bibitem[Chaves \latin{et~al.}(2018)Chaves, Azadani, {\"O}z{\c{c}}elik, Grassi,
  and Low]{chaves2018electrical}
Chaves,~A.; Azadani,~J.; {\"O}z{\c{c}}elik,~V.~O.; Grassi,~R.; Low,~T.
  Electrical control of excitons in van der Waals heterostructures with type-II
  band alignment. \emph{Phys. Rev. B} \textbf{2018}, \emph{98}, 121302\relax
\mciteBstWouldAddEndPuncttrue
\mciteSetBstMidEndSepPunct{\mcitedefaultmidpunct}
{\mcitedefaultendpunct}{\mcitedefaultseppunct}\relax
\EndOfBibitem
\bibitem[Zhuang \latin{et~al.}(1999)Zhuang, Miranda, Kim, and
  Shen]{zhuang1999mapping}
Zhuang,~X.; Miranda,~P.; Kim,~D.; Shen,~Y. Mapping molecular orientation and
  conformation at interfaces by surface nonlinear optics. \emph{Phys. Rev. B}
  \textbf{1999}, \emph{59}, 12632\relax
\mciteBstWouldAddEndPuncttrue
\mciteSetBstMidEndSepPunct{\mcitedefaultmidpunct}
{\mcitedefaultendpunct}{\mcitedefaultseppunct}\relax
\EndOfBibitem
\bibitem[Hore \latin{et~al.}(2005)Hore, Beaman, Parks, and
  Richmond]{hore2005whole}
Hore,~D.~K.; Beaman,~D.~K.; Parks,~D.~H.; Richmond,~G.~L. Whole-molecule
  approach for determining orientation at isotropic surfaces by nonlinear
  vibrational spectroscopy. \emph{J. Phys. Chem. B} \textbf{2005}, \emph{109},
  16846--16851\relax
\mciteBstWouldAddEndPuncttrue
\mciteSetBstMidEndSepPunct{\mcitedefaultmidpunct}
{\mcitedefaultendpunct}{\mcitedefaultseppunct}\relax
\EndOfBibitem
\bibitem[O'Brien and Massari(2015)O'Brien, and Massari]{o2015experimental}
O'Brien,~D.~B.; Massari,~A.~M. Experimental evidence for an optical
  interference model for vibrational sum frequency generation on multilayer
  organic thin film systems. Consideration for higher order terms. \emph{J.
  Chem. Phys.} \textbf{2015}, \emph{142}, 024704\relax
\mciteBstWouldAddEndPuncttrue
\mciteSetBstMidEndSepPunct{\mcitedefaultmidpunct}
{\mcitedefaultendpunct}{\mcitedefaultseppunct}\relax
\EndOfBibitem
\bibitem[Wang \latin{et~al.}(2015)Wang, Velarde, Gan, and
  Fu]{wang2015quantitative}
Wang,~H.-F.; Velarde,~L.; Gan,~W.; Fu,~L. Quantitative sum-frequency generation
  vibrational spectroscopy of molecular surfaces and interfaces: lineshape,
  polarization, and orientation. \emph{Annu. Rev. Phys. Chem.} \textbf{2015},
  \emph{66}, 189--216\relax
\mciteBstWouldAddEndPuncttrue
\mciteSetBstMidEndSepPunct{\mcitedefaultmidpunct}
{\mcitedefaultendpunct}{\mcitedefaultseppunct}\relax
\EndOfBibitem
\bibitem[Li \latin{et~al.}(2015)Li, Wang, and Xiong]{li2015probing}
Li,~Y.; Wang,~J.; Xiong,~W. Probing electronic structures of organic
  semiconductors at buried interfaces by electronic sum frequency generation
  spectroscopy. \emph{J. Phys. Chem. C} \textbf{2015}, \emph{119},
  28083--28089\relax
\mciteBstWouldAddEndPuncttrue
\mciteSetBstMidEndSepPunct{\mcitedefaultmidpunct}
{\mcitedefaultendpunct}{\mcitedefaultseppunct}\relax
\EndOfBibitem
\bibitem[Li \latin{et~al.}(2019)Li, Xiang, and Xiong]{li2019heterodyne}
Li,~Y.; Xiang,~B.; Xiong,~W. Heterodyne transient vibrational SFG to reveal
  molecular responses to interfacial charge transfer. \emph{J. Chem. Phys.}
  \textbf{2019}, \emph{150}, 114706\relax
\mciteBstWouldAddEndPuncttrue
\mciteSetBstMidEndSepPunct{\mcitedefaultmidpunct}
{\mcitedefaultendpunct}{\mcitedefaultseppunct}\relax
\EndOfBibitem
\bibitem[Ohno \latin{et~al.}(2017)Ohno, Wang, and Geiger]{ohno2017second}
Ohno,~P.~E.; Wang,~H.-f.; Geiger,~F.~M. Second-order spectral lineshapes from
  charged interfaces. \emph{Nat. Commun.} \textbf{2017}, \emph{8}, 1032\relax
\mciteBstWouldAddEndPuncttrue
\mciteSetBstMidEndSepPunct{\mcitedefaultmidpunct}
{\mcitedefaultendpunct}{\mcitedefaultseppunct}\relax
\EndOfBibitem
\bibitem[Grimme(2006)]{grimme2006semi}
Grimme,~S. Semiempirical {GGA}-type density functional constructed with a
  long-range dispersion correction. \emph{J. Comput. Chem.} \textbf{2006},
  \emph{27}, 1787--1799\relax
\mciteBstWouldAddEndPuncttrue
\mciteSetBstMidEndSepPunct{\mcitedefaultmidpunct}
{\mcitedefaultendpunct}{\mcitedefaultseppunct}\relax
\EndOfBibitem
\bibitem[Bl\"ochl(1994)]{blochl94}
Bl\"ochl,~P.~E. Projector augmented-wave method. \emph{Phys. Rev. B}
  \textbf{1994}, \emph{50}, 17953--17979\relax
\mciteBstWouldAddEndPuncttrue
\mciteSetBstMidEndSepPunct{\mcitedefaultmidpunct}
{\mcitedefaultendpunct}{\mcitedefaultseppunct}\relax
\EndOfBibitem
\bibitem[Perdew \latin{et~al.}(1996)Perdew, Burke, and Ernzerhof]{pbe}
Perdew,~J.~P.; Burke,~K.; Ernzerhof,~M. Generalized gradient approximation made
  simple. \emph{Phys. Rev. Lett.} \textbf{1996}, \emph{77}, 3865--3868\relax
\mciteBstWouldAddEndPuncttrue
\mciteSetBstMidEndSepPunct{\mcitedefaultmidpunct}
{\mcitedefaultendpunct}{\mcitedefaultseppunct}\relax
\EndOfBibitem
\bibitem[Kresse and Furthm\"uller(1996)Kresse, and Furthm\"uller]{vasp}
Kresse,~G.; Furthm\"uller,~J. Efficient iterative schemes for ab initio
  total-energy calculations using a plane-wave basis set. \emph{Phys. Rev. B}
  \textbf{1996}, \emph{54}, 11169--11186\relax
\mciteBstWouldAddEndPuncttrue
\mciteSetBstMidEndSepPunct{\mcitedefaultmidpunct}
{\mcitedefaultendpunct}{\mcitedefaultseppunct}\relax
\EndOfBibitem
\end{mcitethebibliography}

\end{document}